\documentclass[conference]{IEEEtran}

 \usepackage{mathtools} 
 \usepackage{amsfonts}
 \usepackage{amsmath}
 \usepackage{amsthm}
 \usepackage{amssymb}
 \usepackage{lipsum}
\usepackage{stfloats}
 
\usepackage{blindtext}
\usepackage{amssymb}
\usepackage{amsmath}
\usepackage{graphicx}
\usepackage{cite}
\usepackage{booktabs}
\usepackage{multirow}
\usepackage{bmpsize}
\usepackage{algorithm}
\usepackage{algpseudocode}

\usepackage{booktabs}
\usepackage{enumitem}
\usepackage{subfigure}
\usepackage{graphicx}
\usepackage{xcolor}
\usepackage{url}
\usepackage{cite}  
\usepackage{enumerate}
\usepackage{svg}

 \usepackage{mathtools} 
 \usepackage{amsfonts}
 \usepackage{amsmath}
 \usepackage{amsthm}
 \usepackage{amssymb}

\usepackage{caption}
\usepackage{booktabs}
\usepackage{enumitem}
\usepackage{subfigure}
\usepackage{graphicx}
\usepackage{xcolor}
\usepackage{url}
\usepackage{cite}

\begin{document}

\title{Structured Agentic Learning for Persistent Spatial Acquisition in RDT-TAIS-Assisted Communications}


\author{\IEEEauthorblockA{Bin Liu\IEEEauthorrefmark{1},
Wei~Ni\IEEEauthorrefmark{2},
 Tommy~Svensson\IEEEauthorrefmark{1},
 Sofie Pollin\IEEEauthorrefmark{3}
}

\IEEEauthorblockA{\IEEEauthorrefmark{1} Department of Electrical Engineering, Chalmers University of Technology, Sweden.
\IEEEauthorblockA{\IEEEauthorrefmark{2}School of Engineering, Edith Cowan University, Australia. }
\IEEEauthorblockA{\IEEEauthorrefmark{3}Department of Electrical Engineering, KU Leuven, Belgium.}}
\IEEEauthorblockA{Email: binliu@ieee.org, wei.ni@ieee.org, tommy.svensson@chalmers.se, sofie.pollin@kuleuven.be}
}

\maketitle

\begin{abstract}
Persistent Transparent amplifying intelligent surface (TAIS) operation can reduce repeated radio-information acquisition
by reusing environmental knowledge across control intervals  and store them in a Radio Digital Twin (RDT). This knowledge changes with radio-environment evolution and current communication task.
This paper studies \emph{persistent spatial knowledge maintenance} for TAIS
control and formulates it as a sequential information-acquisition problem
that maximizes effective communication performance while accounting for
physical-acquisition overhead under an outage-reliability constraint.
We develop \emph{structured agentic spatial intelligence (ASI)},
which realizes a communication-oriented agent that maintains persistent
spatial knowledge, forms beliefs from radio and context observations, and
adapts physical information acquisition through closed-loop interaction
with the radio environment.
By jointly exploiting radio-consistent context inference, temporal
knowledge-sufficiency assessment, graph-structured regional reasoning, and
predictive outage-risk feedback, \textit{Structured ASI} determines \emph{when}
additional acquisition is needed, \emph{where} it is directed, and
\emph{how much} physical information is sufficient before stopping or
escalating to full radio-environment reacquisition.
Numerical results show that \textit{Structured ASI} improves the effective rate by
$2.59\%$ while reducing acquisition overhead by $25\%$ at six structural
changes per 100 control intervals, and achieves a $2.82\%$ effective-rate
gain with a $41.8\%$ overhead reduction over a heterogeneous 400-interval
mission.
\end{abstract}


%

\IEEEpeerreviewmaketitle
 \section{Introduction}
\label{sec:introduction}

Transparent amplifying intelligent surfaces (TAISs) provide an engineered
transmission path for indoor-to-outdoor communication across building facades,
where direct propagation can suffer severe penetration loss~\cite{liu2026taismu}.
Previously acquired radio information can also support robust wireless
operation~\cite{KhanColeri2026DTBeam}, but surface configuration still commonly
relies on pilot-based acquisition under prescribed surface states~\cite{chen2023risce}. Repeated probing can therefore consume substantial radio
resources as the surface grows.

Wireless propagation exhibits persistent temporal and spatial structure that
can be reused across communication intervals~\cite{zeng2021toward,cao2025channeltwinning}.
In indoor-to-outdoor deployments, building geometry and dominant propagation
structures typically evolve more slowly than the instantaneous channel, which
creates an opportunity to reduce repeated acquisition by maintaining spatial
knowledge of the radio environment.

The usefulness of maintained spatial knowledge nevertheless changes with both
the physical environment and the communication task. Structural changes can
invalidate maintained information, whereas user movement or a context change
can alter which regions influence the current TAIS-assisted link. Hence,
outdated information can remain task-irrelevant, while physically valid
information can become newly relevant. Effective knowledge reuse must determine
\emph{when} the maintained knowledge is insufficient~\cite{Uzlaner2025Drift,KhanColeri2024EventTriggered}, \emph{where} fresh
information is required~\cite{LiMao2024DynamicGNN}, and \emph{how much}
physical acquisition is sufficient~\cite{Zhang2024RISActiveLocalization}.
These coupled decisions form a \emph{WHEN--WHERE--HOW MUCH} sequential
information-acquisition problem.

Existing work addresses these components separately. RIS acquisition reduces
the pilot burden for surface configuration~\cite{Zhang2024RISActiveLocalization},
while channel knowledge maps, channel twinning, and active radio-map
reconstruction support persistent propagation-knowledge reuse and selective
updating~\cite{polyzos2024bayesian}. Learning-based wireless control has
incorporated event-triggered decisions and explicit reliability or safety
requirements~\cite{ReyhanColeri2025SafeDRL}; sequential change-point methods
detect persistent changes~\cite{fearnhead2007online}; and discrepancy-driven
radio-map updating provides a residual-triggered maintenance principle~\cite{yang2025ckm}. However, these approaches do not jointly determine which
maintained spatial information is relevant to the current task and how much
fresh acquisition is sufficient under a communication-level reliability
requirement.

This paper formulates persistent spatial knowledge maintenance for long-term
TAIS operation as a sequential information-acquisition problem that maximizes
effective communication performance while accounting for physical-acquisition
overhead under an outage-reliability constraint. We develop \emph{Structured
Agentic Spatial Intelligence (Structured ASI)} as a communication-oriented
acquisition policy that maintains persistent spatial knowledge  in a Radio Digital Twin (RDT), forms
controller-visible beliefs from routine radio and context observations, and
adapts acquisition through closed-loop interaction with the radio environment~\cite{Zheng2026AgenticDRL}.

The main contributions are summarized as follows.
\begin{itemize}
\item
We establish a multi-timescale radio-environment model that separates
persistent regional states, structural evolution, transient propagation
variation, and task-dependent spatial relevance, and formulate persistent
spatial knowledge maintenance as a causal sequential information-acquisition
problem under an outage-reliability requirement.

\item
We develop Structured ASI with radio-consistent context inference, temporal
knowledge-sufficiency assessment, and graph-structured regional reasoning to
determine when additional acquisition is required and where fresh information
should be obtained for the RDT.

\item
We develop a reliability-driven verification--reacquisition policy that
updates controller information after each physical observation, stops regional
acquisition once the predictive-reliability target is met, and escalates to
full radio-environment reacquisition when regional recovery remains
insufficient.
\end{itemize}

Numerical results show that Structured ASI improves the effective rate by
$2.59\%$ while reducing acquisition overhead by $25\%$ at six structural
changes per 100 control intervals, and achieves a $2.82\%$ effective-rate gain
with a $41.8\%$ overhead reduction over a heterogeneous 400-interval mission.

 \section{System Model and Problem Formulation}
\label{sec:system_model}

\begin{figure}[t]
    \centering
    \includegraphics[width=\columnwidth]{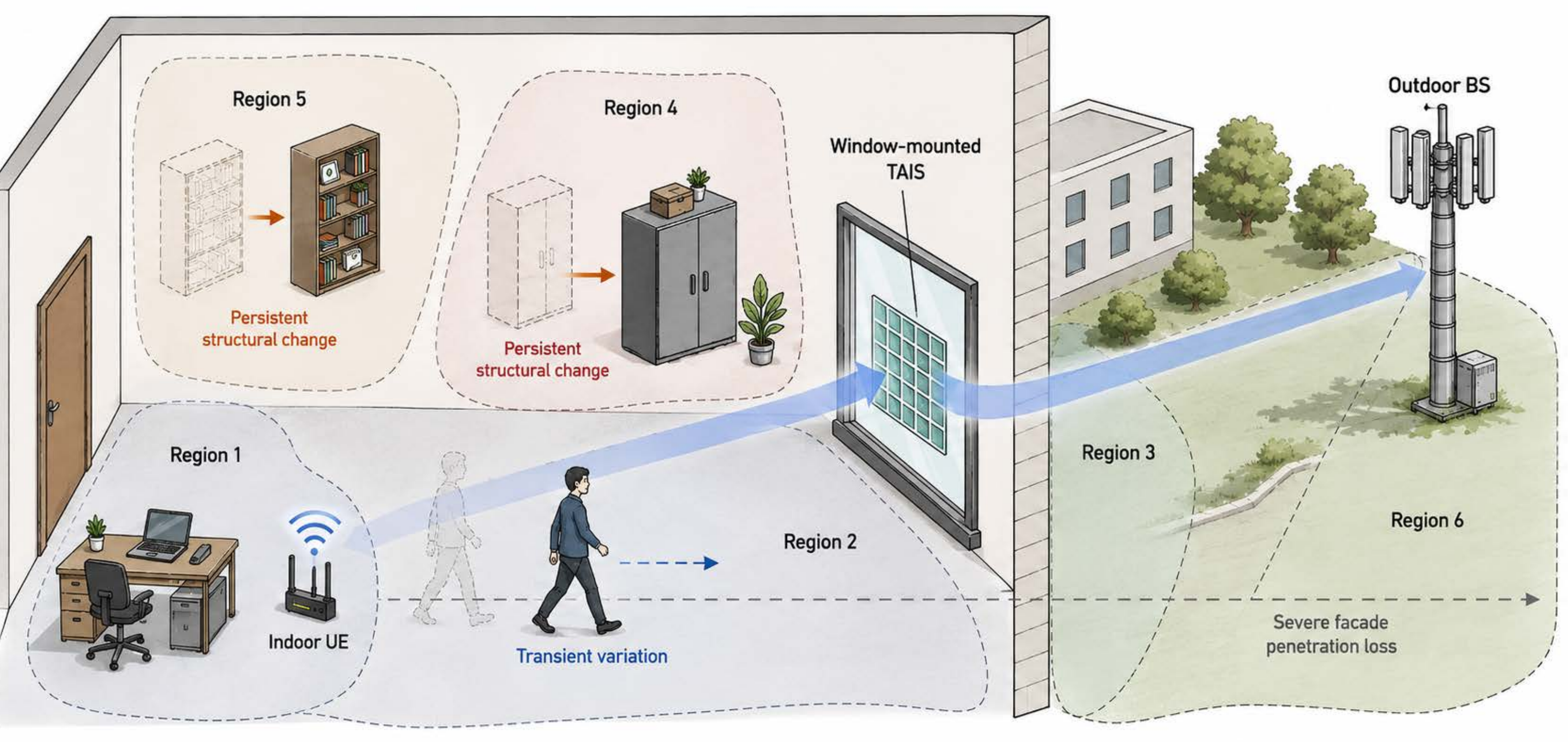}
    \caption{TAIS-assisted indoor-to-outdoor communication with persistent structural changes, transient propagation variation, and task-relevant spatial regions.}
    \label{fig:scenario}
    \vspace{-3mm}
\end{figure}

We consider the TAIS-assisted indoor-to-outdoor uplink in
Fig.~\ref{fig:scenario}, where a single-antenna indoor user equipment
(UE) communicates with an $N_{\rm BS}$-antenna outdoor BS through a
window-mounted TAIS with $M$ active transmissive elements. Because
facade penetration loss can severely attenuate the direct link, the
engineered UE--TAIS--BS path serves as the primary communication
link~\cite{liu2026taismu}.

During control interval $t$, the TAIS applies the payload configuration
$\boldsymbol{\phi}_t=[\phi_{1,t},\ldots,\phi_{M,t}]^{\mathsf T}\in\mathcal A$,
where $\mathcal A$ is the hardware-feasible set and
$\mathbf\Phi_t=\operatorname{diag}(\boldsymbol{\phi}_t)$. The received
uplink signal is
\begin{equation}
\mathbf y_t^{\rm com}
=
\sqrt{P_u}\,
\mathbf F_t\mathbf\Phi_t\mathbf h_t x_t
+
\mathbf n_t,
\label{eq:received_signal}
\end{equation}
where $\mathbf h_t\in\mathbb C^M$ and
$\mathbf F_t\in\mathbb C^{N_{\rm BS}\times M}$ denote the UE--TAIS and
TAIS--BS channels, respectively, $P_u$ is the UE transmit power, $x_t$
is the unit-power payload symbol with $\mathbb E[|x_t|^2]=1$, and
$\mathbf n_t\sim\mathcal{CN}(\mathbf0,\sigma_0^2\mathbf I)$ is the
receiver noise.

With maximal-ratio combining at the BS, the payload rate associated with
\eqref{eq:received_signal} is
\begin{equation}
R_t
=
\log_2
\left(
1+
\frac{
P_u
\left\|
\mathbf F_t\mathbf\Phi_t\mathbf h_t
\right\|^2
}{
\sigma_0^2
}
\right).
\label{eq:payload_rate}
\end{equation}
\subsection{Persistent Radio Environment and Controller Observations}
\label{subsec:environment}

We model the reusable radio-environment structure through $K$
persistent spatial regions $\mathcal V=\{1,\ldots,K\}$ connected by the
graph $\mathcal G=(\mathcal V,\mathcal E)$
~\cite{zeng2021toward,cao2025channeltwinning}. Each region $k$ is
represented in a common state space by a static descriptor $\mathbf p_k$
and a persistent regional state $\mathbf s_{k,t}$. Collecting the regional
states in $\mathbf s_t$, their evolution is modeled as
\begin{equation}
\mathbf s_t
=
f_{\rm env}
\left(
\mathbf s_{t-1},
\boldsymbol{\nu}_t,
\boldsymbol{\delta}_t
\right),
\quad
\mathcal D_t
=
\left\{
k\in\mathcal V:
\boldsymbol{\delta}_{k,t}\neq\mathbf0
\right\},
\label{eq:environment_evolution}
\end{equation}
where $\boldsymbol{\nu}_t$ represents slow persistent evolution,
$\boldsymbol{\delta}_t=\{\boldsymbol{\delta}_{k,t}\}_{k\in\mathcal V}$
represents structurally significant regional changes, and
$\mathcal D_t$ is the corresponding structural-change support. In
addition, $\boldsymbol{\xi}_t$ denotes transient propagation variation
that perturbs the instantaneous radio condition without changing the
persistent state $\mathbf s_t$. Hence,
$\boldsymbol{\delta}_t$ changes the validity of maintained spatial
knowledge, whereas $\boldsymbol{\xi}_t$ does not.

The current communication task is represented by the context
$\chi_t\in\mathcal C$. For context $\chi$, the normalized regional
contribution profile
$\mathbf w(\chi)=[w_1(\chi),\ldots,w_K(\chi)]^{\mathsf T}$ satisfies
$w_k(\chi)\geq0$ and $\sum_k w_k(\chi)=1$. Its nonzero support,
\begin{equation}
\mathcal V^{\rm rel}(\chi)
=
\left\{
k\in\mathcal V:
w_k(\chi)>0
\right\},
\label{eq:relevant_support}
\end{equation}
contains the regions relevant to that communication context. In general,
the structural-change support $\mathcal D_t$ and the task-relevant
support $\mathcal V^{\rm rel}(\chi_t)$ need not coincide. This distinction
is the basis for acquiring fresh information selectively rather than
refreshing every changed region.

Conditioned on the persistent radio-environment state $\mathbf s_t$,
communication context $\chi_t$, and transient variation
$\boldsymbol{\xi}_t$, the instantaneous channels follow
\begin{equation}
(\mathbf h_t,\mathbf F_t)
\sim
p_{\rm ch}
\left(
\mathbf h,\mathbf F
\mid
\mathbf s_t,\chi_t,\boldsymbol{\xi}_t
\right).
\label{eq:channel_law}
\end{equation}
The controller does not observe these latent quantities directly. During
normal operation, it receives the low-overhead routine radio observation
$\mathbf r_t$ and the external context observation
$\widetilde{\chi}_t$:
\begin{subequations}
\label{eq:routine_observations}
\begin{align}
\mathbf r_t
&=
g_r
\left(
\mathbf h_t,\mathbf F_t,\boldsymbol{\phi}_{t-1}
\right)
+
\mathbf n_t^r,
\\
\widetilde{\chi}_t
&\sim
p_{\rm ext}
\left(
\widetilde{\chi}\mid\chi_{t-d_t}
\right),
\quad d_t\geq0.
\end{align}
\end{subequations}
Here, $g_r(\cdot)$ is the routine-observation mapping,
$\mathbf n_t^r$ is the corresponding observation noise, and $d_t$ is the
context-reporting delay. The acquisition policy must therefore infer
whether a change in $\mathbf r_t$ reflects stale persistent knowledge,
transient variation $\boldsymbol{\xi}_t$, or delayed context information.

\subsection{Physical Acquisition and Persistent Spatial Knowledge}
\label{subsec:acquisition_model}

The controller obtains additional physical information through
\emph{regional verification}, \emph{regional reacquisition}, or
\emph{full radio-environment reacquisition}. Regional verification tests
the consistency of maintained information, whereas regional
reacquisition obtains fresh information for a selected region. Let
$\Omega_t^{\rm ver}\subseteq\mathcal V$ denote the verified regions and
$\Omega_t^{\rm reacq}\subseteq\Omega_t^{\rm ver}$ those subsequently
reacquired. Their physical observations are modeled as
\begin{subequations}
\label{eq:physical_acquisition}
\begin{align}
\mathbf y_{k,t}^{\rm ver}
&=
g_k^{\rm ver}(\mathbf s_{k,t})
+
\mathbf n_{k,t}^{\rm ver},
\quad
k\in\Omega_t^{\rm ver},
\\
\mathbf y_{k,t}^{\rm reacq}
&=
g_k^{\rm reacq}(\mathbf s_{k,t})
+
\mathbf n_{k,t}^{\rm reacq},
\quad
k\in\Omega_t^{\rm reacq}.
\end{align}
\end{subequations}
Here, $g_k^{\rm ver}(\cdot)$ and $g_k^{\rm reacq}(\cdot)$ are the
regional verification and regional reacquisition observation mappings,
respectively, while $\mathbf n_{k,t}^{\rm ver}$ and
$\mathbf n_{k,t}^{\rm reacq}$ are the corresponding acquisition noises.
Full radio-environment reacquisition obtains fresh regional information
for all $k\in\mathcal V$; its binary decision is denoted by
$b_t^{\rm full}\in\{0,1\}$.

Each physical action consumes radio resources. The acquisition overhead
in interval $t$ and over the complete horizon are
\begin{equation}
\begin{aligned}
N_t^{\rm acq}
&=
\sum_{k\in\Omega_t^{\rm ver}}
N_{k,t}^{\rm ver}
+
\sum_{k\in\Omega_t^{\rm reacq}}
N_{k,t}^{\rm reacq}
+
b_t^{\rm full}N_t^{\rm full},
\\
N^{\rm acq}
&=
\sum_{t=1}^{T}N_t^{\rm acq}.
\end{aligned}
\label{eq:acquisition_cost}
\end{equation}
where $N_{k,t}^{\rm ver}$, $N_{k,t}^{\rm reacq}$, and $N_t^{\rm full}$
are the symbol costs of regional verification, regional reacquisition,
and full radio-environment reacquisition, respectively.

The acquired information is retained across control intervals in the
persistent spatial memory
\begin{equation}
\mathcal M_t
=
\left\{
\widehat{\mathbf s}_{k,t},
\mathbf\Sigma_{k,t},
a_{k,t}
\right\}_{k\in\mathcal V},
\label{eq:persistent_memory}
\end{equation}
where $\widehat{\mathbf s}_{k,t}$ is the maintained regional-state
estimate, $\mathbf\Sigma_{k,t}\succeq\mathbf0$ quantifies its
uncertainty, and $a_{k,t}$ is the number of control intervals elapsed
since the most recent regional or full reacquisition of region $k$.
Regional verification can tighten the uncertainty while retaining the
maintained state, regional reacquisition refreshes the corresponding
state and uncertainty, and unacquired information is retained for later
reuse.

At the beginning of interval $t$, these retained quantities are combined
with the current routine observations and the preceding link outcome in
the controller-visible pre-acquisition information state
\begin{equation}
\mathcal I_t^-
=
\left\{
\mathcal M_{t-1},
\mathbf r_t,
\widetilde{\chi}_t,
\boldsymbol{\phi}_{t-1},
R_{t-1}
\right\}.
\label{eq:pre_acquisition_information}
\end{equation}
An initialized memory $\mathcal M_0$, a feasible configuration
$\boldsymbol{\phi}_0\in\mathcal A$, and its corresponding initial link
rate $R_0$ provide the starting information for $t=1$.

\subsection{Sequential Acquisition Problem}
\label{subsec:problem}

We evaluate the acquisition policy through the effective rate
$R_{\rm eff}$ and the horizon-averaged outage probability $P_{\rm out}$,
which capture acquisition-adjusted communication performance and
reliability, respectively:
\begin{subequations}
\label{eq:performance_metrics}
\begin{align}
R_{\rm eff}
&=
\frac{
N_{\rm sym}
}{
TN_{\rm sym}+N^{\rm acq}
}
\sum_{t=1}^{T}R_t,
\\
P_{\rm out}
&=
\frac{1}{T}
\sum_{t=1}^{T}
\Pr
\left(
R_t<R_{\rm th}
\right),
\end{align}
\end{subequations}
where $N_{\rm sym}$ is the number of payload OFDM symbols per control
interval and $R_{\rm th}$ is the prescribed rate target.

The acquisition decisions are represented by the causal policy
$\pi_{\rm acq}=(\pi_{\rm loc},\pi_{\rm reacq},\pi_{\rm full})$. Starting
from $\mathcal I_{t,0}=\mathcal I_t^-$, $\pi_{\rm loc}$ selects a region
$k_{t,\ell}\in\mathcal V$ for verification or stops the regional
acquisition process. After each verification,
$\pi_{\rm reacq}$ decides whether the selected region is reacquired, and
each resulting physical observation updates $\mathcal I_{t,\ell}$ before
the next decision. After regional acquisition terminates,
$\pi_{\rm full}$ decides whether full radio-environment reacquisition is
required. The resulting post-acquisition information state
$\mathcal I_t^+$ determines the payload configuration through the
prescribed TAIS-control mapping
$
\boldsymbol{\phi}_t
=
\pi_{\rm TAIS}
\left(
\mathcal I_t^+
\right)$,
$\boldsymbol{\phi}_t\in\mathcal A$.
The prescribed mapping $\pi_{\rm TAIS}$ is shared by all acquisition
policies and is not optimized in the knowledge-maintenance problem.

The persistent spatial knowledge-maintenance problem is therefore to
maximize acquisition-adjusted communication performance while satisfying
the horizon-level reliability requirement:
\begin{subequations}
\label{prob:knowledge_maintenance}
\begin{align}
\mathbf{P1:}\quad
\max_{\pi_{\rm acq}\in\Pi_{\rm acq}^{\rm causal}}
\quad&
\mathbb E_{\pi_{\rm acq}}
\left[
R_{\rm eff}
\right]
\\
\mathrm{s.t.}\quad&
P_{\rm out}(\pi_{\rm acq})
\leq
\epsilon,
\end{align}
\end{subequations}
where $\Pi_{\rm acq}^{\rm causal}$ contains policies that use only the
controller-visible information available at each decision step, and
$\epsilon$ is the prescribed horizon-level outage tolerance.

 \section{Structured ASI for Persistent TAIS Control}
\label{sec:asi}

\begin{figure}[!t]
    \centering
    \includegraphics[width=0.65\columnwidth]{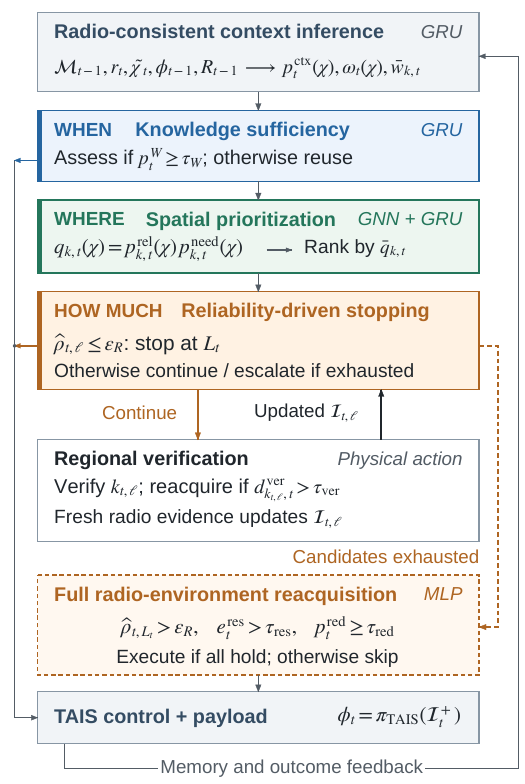}
    \caption{Structured ASI realization of the \emph{WHEN--WHERE--HOW MUCH} acquisition process.}
    \label{fig:structured_asi}
    \vspace{-3mm}
\end{figure}

Problem~\eqref{prob:knowledge_maintenance} requires the controller to
determine whether additional information is needed, where it should be
acquired, and when acquisition can stop. We develop Structured ASI to
realize these coupled decisions from the controller-visible information
state $\mathcal I_t^-$ through a \emph{WHEN--WHERE--HOW MUCH}
verification--reacquisition process. As illustrated in
Fig.~\ref{fig:structured_asi}, radio-consistent context inference and
temporal knowledge-sufficiency assessment determine \emph{when}
additional acquisition is required, graph-structured regional reasoning
determines \emph{where} fresh information should be acquired, and
predictive outage-risk feedback determines \emph{how much} acquisition
is sufficient and whether full radio-environment reacquisition is
required.

\subsection{WHEN: Controller-Visible Knowledge Sufficiency}
\label{subsec:when}

The external context observation $\widetilde{\chi}_t$ can be delayed or
inconsistent with the current radio condition. Structured ASI therefore
reconciles it with the routine radio observation $\mathbf r_t$ before
using context information for spatial acquisition. For each context
hypothesis $\chi\in\mathcal C$, the radio-consistency statistic is
\begin{equation}
e_t^{\rm ctx}(\chi)
=
d_r
\left(
\mathbf r_t,
\widehat{\mathbf r}_t
\left(
\mathcal M_{t-1},
\boldsymbol{\phi}_{t-1},
\chi
\right)
\right),
\label{eq:context_consistency}
\end{equation}
where
$\widehat{\mathbf r}_t(\mathcal M_{t-1},\boldsymbol{\phi}_{t-1},\chi)$
is the routine radio response predicted from the maintained spatial memory
under context $\chi$ and the deployed observation configuration
$\boldsymbol{\phi}_{t-1}$, and $d_r(\cdot,\cdot)$ is a
radio-discrepancy measure~\cite{Uzlaner2025Drift}.

A GRU followed by a softmax prediction head processes the
controller-visible quantities in $\mathcal I_t^-$ together with the
context-conditioned consistency statistics
$\{e_t^{\rm ctx}(\chi)\}_{\chi\in\mathcal C}$ and produces the
normalized context belief $p_t^{\rm ctx}(\chi)$. Let
\begin{equation}
    \arg\max_{\chi\in\mathcal C}
p_t^{\rm ctx}(\chi).
\end{equation}
The context weight used by the subsequent spatial prioritization and
reliability evaluation is
\begin{equation}
\omega_t(\chi)
=
\begin{cases}
\mathbf 1\{\chi=\widetilde{\chi}_t\},
&
\widetilde{\chi}_t=\chi_t^{\rm MAP},
\\
p_t^{\rm ctx}(\chi),
&
\widetilde{\chi}_t\neq\chi_t^{\rm MAP},
\end{cases}
\label{eq:context_weight}
\end{equation}
which yields the controller-visible regional contribution weight
\begin{equation}
\bar w_{k,t}
=
\sum_{\chi\in\mathcal C}
\omega_t(\chi)w_k(\chi).
\label{eq:regional_weight}
\end{equation}

A second GRU evaluates whether the maintained spatial knowledge remains
sufficient by processing $\mathcal I_t^-$, the context belief, and the
radio-consistency evidence. Its output
$p_t^{\rm W}\in[0,1]$ denotes the knowledge-insufficiency belief.
Regional acquisition is activated only when
\begin{equation}
p_t^{\rm W}\geq\tau_{\rm W},
\label{eq:when_trigger}
\end{equation}
and the maintained knowledge is otherwise reused. This event-triggered
structure avoids repeated physical acquisition when the available
knowledge remains sufficient~\cite{KhanColeri2024EventTriggered}.

\subsection{WHERE: Task-Relevant Spatial Prioritization}
\label{subsec:where}

Once acquisition is activated, Structured ASI identifies regions for
which fresh information is relevant to the current communication task
and the maintained information is likely to require an update. A GNN processes the
maintained regional state $\widehat{\mathbf s}_{k,t-1}$, knowledge age
$a_{k,t-1}$, and static descriptor $\mathbf p_k$ over the spatial graph
$\mathcal G$, while a context-conditioned GRU captures the temporal
evolution of the maintained regional information. The resulting spatio-temporal representation produces the
task-relevance belief $p_{k,t}^{\rm rel}(\chi)$ and update-need belief
$p_{k,t}^{\rm need}(\chi)$~\cite{LiMao2024DynamicGNN}.

The regional acquisition priority is factorized as
\begin{equation}
q_{k,t}(\chi)
=
p_{k,t}^{\rm rel}(\chi)
p_{k,t}^{\rm need}(\chi),
\label{eq:priority_score}
\end{equation}
so that a region receives high priority only when both task relevance
and update need are high. Context uncertainty is
incorporated through
\begin{subequations}
\label{eq:context_marginal_priority}
\begin{align}
\bar p_{k,t}^{\rm rel}
&=
\sum_{\chi\in\mathcal C}
\omega_t(\chi)
p_{k,t}^{\rm rel}(\chi),
\\
\bar q_{k,t}
&=
\sum_{\chi\in\mathcal C}
\omega_t(\chi)
q_{k,t}(\chi).
\end{align}
\end{subequations}
The resulting candidate region set is
\begin{equation}
\mathcal V_t^{\rm cand}
=
\left\{
k\in\mathcal V:
\bar p_{k,t}^{\rm rel}\geq\tau_{\rm rel}
\right\},
\label{eq:candidate_set}
\end{equation}
and its regions are examined in descending order of $\bar q_{k,t}$.
If the thresholded set is empty, the region with the largest
$\bar p_{k,t}^{\rm rel}$ is retained as the fallback candidate.

\begin{figure*}[t]
    \centering
    \includegraphics[width=\textwidth]
{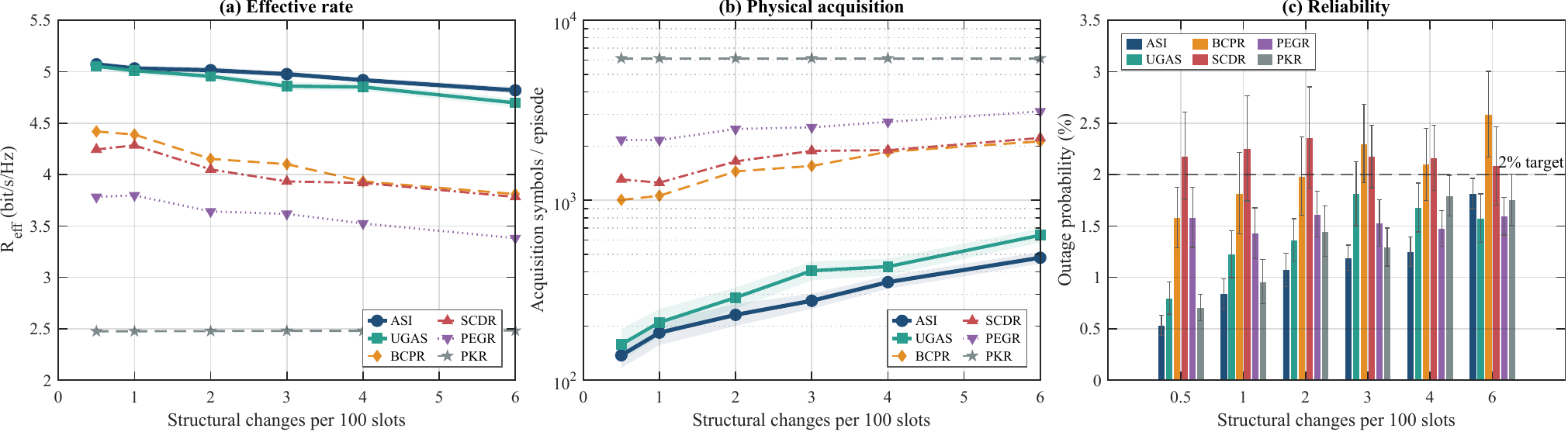}
    \caption{\small Robustness to structural-change frequency:
    (a) effective rate,
    (b) physical-acquisition overhead, and
    (c) outage probability versus the number of persistent structural changes
    per 100 control intervals. The horizontal dashed line in (c) denotes the
    outage tolerance $\epsilon=2\%$.}
    \label{fig:structural_rate}
\end{figure*}

\subsection{HOW MUCH: Closed-Loop Acquisition and Stopping}
\label{subsec:howmuch}

For the $\ell$-th selected region $k_{t,\ell}$, Structured ASI compares
the verification observation with the response predicted from the
current controller information state:
\begin{subequations}
\label{eq:regional_decision}
\begin{align}
d_{k_{t,\ell},t}^{\rm ver}
&=
d_{\rm ver}
\left(
\mathbf y_{k_{t,\ell},t}^{\rm ver},
\widehat{\mathbf y}_{k_{t,\ell},t}^{\rm ver}
(\mathcal I_{t,\ell-1})
\right),
\\
b_{t,\ell}^{\rm reacq}
&=
\mathbf 1
\left\{
d_{k_{t,\ell},t}^{\rm ver}
>
\tau_{\rm ver}
\right\}.
\end{align}
\end{subequations}
Here, $\widehat{\mathbf y}_{k_{t,\ell},t}^{\rm ver}(\mathcal I_{t,\ell-1})$
denotes the predicted verification response, and
$d_{\rm ver}(\cdot,\cdot)$ is the corresponding verification-discrepancy measure.
The sequential stage is initialized with
$\widehat{\mathbf s}_{k,t,0}=\widehat{\mathbf s}_{k,t-1}$ and
$\mathbf\Sigma_{k,t,0}=\mathbf\Sigma_{k,t-1}$. A negative verification
retains the regional-state estimate while refining its uncertainty,
whereas $b_{t,\ell}^{\rm reacq}=1$ refreshes the selected regional
state and uncertainty using the reacquisition observation. Unselected
regions remain unchanged. Each new physical observation is incorporated
into $\mathcal I_{t,\ell}$ before the next acquisition decision.

The acquisition scope is controlled by predictive communication risk.
For acquisition step $\ell$, define the hypothetical payload configuration
\begin{equation}
\boldsymbol{\phi}_{t,\ell}
=
\pi_{\rm TAIS}
\left(
\mathcal I_{t,\ell}
\right)
\label{eq:candidate_tais}
\end{equation}
as the TAIS configuration that would be used if regional acquisition
terminated at the current information state. The maintained regional
information is then aggregated according to the current contribution
weights $\bar w_{k,t}$:
\begin{align}  \label{eq:aggregate_knowledge}
\widehat{\mathbf s}_{t,\ell}^{\rm agg}
&=
\sum_{k\in\mathcal V}
\bar w_{k,t}
\widehat{\mathbf s}_{k,t,\ell},
\quad
\mathbf\Sigma_{t,\ell}^{\rm agg}
\approx
\sum_{k\in\mathcal V}
\bar w_{k,t}^{\,2}
\mathbf\Sigma_{k,t,\ell}.
\end{align}
The approximation in \eqref{eq:aggregate_knowledge} neglects
cross-region estimation-error covariances.

A local conditional rate model maps the aggregate spatial belief and
the candidate configuration $\boldsymbol{\phi}_{t,\ell}$ to the
predictive payload-rate distribution, while retaining the preceding
realized rate $R_{t-1}$ through $\mathcal I_{t,\ell}$. Its first two
moments are
\begin{equation}
\mu_{t,\ell}
\triangleq
\mathbb E
\left[
R_t
\mid
\mathcal I_{t,\ell},
\boldsymbol{\phi}_{t,\ell}
\right],
\
v_{t,\ell}
\triangleq
\operatorname{Var}
\left(
R_t
\mid
\mathcal I_{t,\ell},
\boldsymbol{\phi}_{t,\ell}
\right).
\label{eq:predictive_moments}
\end{equation}
Under the local Gaussian approximation
$R_t\mid\mathcal I_{t,\ell},\boldsymbol{\phi}_{t,\ell}\approx
\mathcal N(\mu_{t,\ell},v_{t,\ell})$, the predictive outage risk is
\begin{equation}
\widehat{\rho}_{t,\ell}
=
\Psi
\left(
\frac{
R_{\rm th}-\mu_{t,\ell}
}{
\sqrt{v_{t,\ell}}
}
\right),
\label{eq:predictive_risk}
\end{equation}
where $\Psi(\cdot)$ is the standard Gaussian cumulative distribution
function.

If $p_t^{\rm W}<\tau_{\rm W}$, no regional acquisition is performed and
$L_t=0$. Otherwise, the regional acquisition length is
\begin{equation}
L_t
=
\min
\left(
\left\{
\ell\in
\{0,\ldots,|\mathcal V_t^{\rm cand}|\}:
\widehat{\rho}_{t,\ell}
\leq
\epsilon_R
\right\}
\cup
\left\{
|\mathcal V_t^{\rm cand}|
\right\}
\right).
\label{eq:stopping_rule}
\end{equation}
Hence, regional acquisition continues in descending order of
$\bar q_{k,t}$ only while the predictive outage risk exceeds
$\epsilon_R$, and terminates once the predictive outage-risk target is
met or all candidate regions have been examined. The online
threshold $\epsilon_R$ controls within-interval stopping, whereas
$\epsilon$ in Problem~\ref{prob:knowledge_maintenance} specifies the
horizon-level outage requirement. The value of $\epsilon_R$ is selected
during offline calibration together with the remaining operating
thresholds.

\subsection{Residual Escalation and Offline Learning}
\label{subsec:full_reacquisition}

After regional acquisition terminates, Structured ASI evaluates whether
broader radio-environment information is still required. The residual
radio discrepancy is
\begin{equation}
e_t^{\rm res}
=
d_r
\left(
\mathbf r_t,
\widehat{\mathbf r}_{t,L_t}
\right),
\label{eq:residual_discrepancy}
\end{equation}
where $\widehat{\mathbf r}_{t,L_t}$ is the routine-radio response
predicted from the terminal regional information state under the same
observation configuration $\boldsymbol{\phi}_{t-1}$ as $\mathbf r_t$.
An MLP combines
$e_t^{\rm res}$ with the terminal predictive risk, context and regional
beliefs, and the regional acquisition outcome to produce the
spatial-reducibility belief $p_t^{\rm red}$. This belief represents
whether unresolved task-relevant stale spatial information remains and
can be refreshed through full radio-environment reacquisition.

The full-reacquisition decision is
\begin{equation}
b_t^{\rm full}
=
\mathbf 1
\left\{
p_t^{\rm W}\geq\tau_{\rm W},\;
\widehat{\rho}_{t,L_t}>\epsilon_R,\;
e_t^{\rm res}>\tau_{\rm res},\;
p_t^{\rm red}\geq\tau_{\rm red}
\right\}.
\label{eq:full_reacquisition}
\end{equation}
Full reacquisition is therefore activated only when acquisition is
triggered, the terminal predictive risk remains above $\epsilon_R$, the
routine radio evidence remains inconsistent with the maintained knowledge,
and the residual mismatch is judged spatially reducible.

The GRU, GNN, and MLP components are trained offline from
radio-environment trajectories using physical information available only
during training to supervise the context, knowledge-insufficiency,
task-relevance, update-need, and spatial-reducibility beliefs. Their learned parameters
remain fixed during closed-loop evaluation, while the operating
thresholds are selected on a disjoint calibration set. Online
adaptation occurs only through updates of the persistent spatial memory
$\mathcal M_t$ and the controller-visible information state.
The terminal regional or full reacquisition outcome updates
$\mathcal M_t$ and yields the post-acquisition information state
$\mathcal I_t^+$, from which the payload configuration
$\boldsymbol{\phi}_t=\pi_{\rm TAIS}(\mathcal I_t^+)$ is generated.
 \section{Numerical Results}
\label{sec:results}

This section evaluates Structured ASI under persistent structural changes,
transient propagation variation, and imperfect routine radio observations.
The default setup uses $K=16$ spatial regions, four communication contexts,
$M=64$ TAIS elements, and $T=400$ control intervals, with
$N_{\rm sym}=14$ and $R_{\rm th}=5$ bit/s/Hz. Unless otherwise stated,
persistent structural changes occur twice per 100 control intervals, with 30
transient propagation events per episode and acquisition assessment every four
control intervals. Regional verification, regional reacquisition, and full
radio-environment reacquisition cost 2, 24, and 384 symbols, respectively.
All learned parameters are frozen before closed-loop evaluation, the operating
thresholds are determined on a disjoint calibration set, and
$\epsilon_R=\epsilon=0.02$. All schemes use matched environmental
realizations and the same TAIS-control mapping, observation interface, and
acquisition-overhead accounting.

We compare Structured ASI with five adapted reference policies:
\emph{Uncertainty-Guided Active Sensing (UGAS)} verifies regions with high
predictive uncertainty~\cite{polyzos2024bayesian};
\emph{Bayesian Change-Point Reacquisition (BCPR)} triggers full reacquisition
from sequential Bayesian change evidence~\cite{fearnhead2007online};
\emph{Sequential Change-Detection Refresh (SCDR)} uses a CuSum-style
persistent-mismatch statistic~\cite{banerjee2013data};
\emph{Prediction-Error-Gated Refresh (PEGR)} triggers full reacquisition from
a calibrated radio-prediction discrepancy~\cite{yang2025ckm}; and
\emph{Periodic Knowledge-Map Refresh (PKR)} performs fixed-period full
reacquisition~\cite{zeng2021toward}.

\begin{figure}[!t]
    \centering
    \includegraphics[width=0.43\textwidth]
    {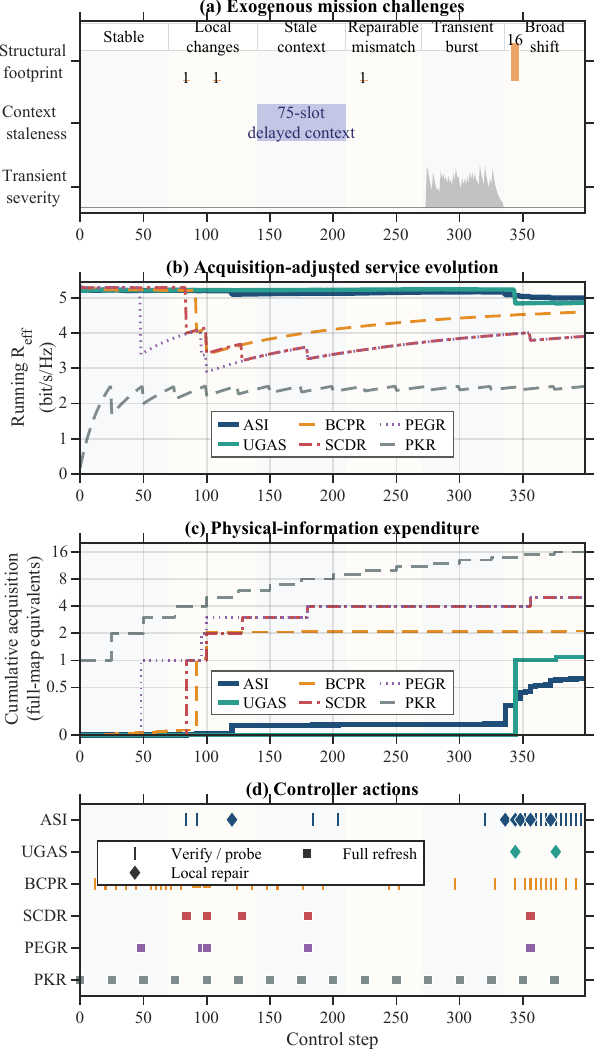}
    \caption{\small Mission-level behavior of the six controllers under the same
    exogenous event sequence.
    (a) Transient propagation variation $\xi_t$, external-context delay
    $d_t$, and structural-change support size $|\mathcal D_t|$;
    (b) running acquisition-adjusted effective rate;
    (c) cumulative acquisition overhead in full-reacquisition equivalents; 
    (d) acquisition actions.
    Numbers above the structural changes in (a) indicate the number of changed
    spatial regions.}
    \label{fig:mission}
    \vspace{-5mm}
\end{figure}

Fig.~\ref{fig:structural_rate} compares $R_{\rm eff}$, physical-acquisition
overhead $N^{\rm acq}$, and outage probability $P_{\rm out}$ as persistent
structural changes become more frequent. At six structural changes per 100
control intervals, Structured ASI improves $R_{\rm eff}$ by $2.59\%$ and
reduces $N^{\rm acq}$ by $25\%$ relative to UGAS, while maintaining
$P_{\rm out}=1.815\%<\epsilon=2\%$. The result reflects task-relevant spatial
prioritization and reliability-driven stopping, which avoid uniform
reacquisition as persistent knowledge becomes outdated more frequently.

Fig.~\ref{fig:mission} evaluates the controllers over a heterogeneous
400-interval mission combining localized structural changes, delayed context
observations, transient propagation variation, and a final broad structural
change. Over the complete mission, Structured ASI achieves $4.995$ bit/s/Hz
with 242 acquisition symbols, corresponding to a $2.82\%$ effective-rate gain
and a $41.8\%$ acquisition reduction relative to UGAS. It performs 37 regional
verifications and 7 regional reacquisitions without full radio-environment
reacquisition Fig.~\ref{fig:mission}(d), illustrating persistent knowledge
reuse during stable intervals and selective repair after detected regional
inconsistency.

\section{Conclusion}

This paper cast persistent spatial knowledge maintenance for
RDT-TAIS-assisted communications as a sequential information-acquisition
problem under communication and reliability constraints. We developed
Structured ASI, which combines persistent spatial knowledge,
task-relevant regional reasoning, verification and
reacquisition, and predictive outage-risk feedback to adapt acquisition
timing, regional focus, and acquisition scope through a
WHEN--WHERE--HOW MUCH process. Numerical results showed that Structured
ASI improves the effective rate by $2.59\%$ while reducing acquisition
overhead by $25\%$ at six structural changes per 100 control
intervals. Over a heterogeneous 400-interval mission, it achieves a
$2.82\%$ effective-rate gain with a $41.8\%$ reduction in acquisition
overhead while maintaining the required reliability.

\section*{acknowledgements}
This work was supported by the Swedish Research Council (VR) through the 6G-NTN-E Research Environment under Grant 2024-06645

\bibliographystyle{IEEEtran_pages_before_year}
\bibliography{mybib}

\appendices

\ifCLASSOPTIONcaptionsoff
  \newpage
\fi

\end{document}